\documentclass[%
aip,
amsmath,amssymb,
reprint,%
]{revtex4-1}

\usepackage{graphicx}% Include figure files
\usepackage{dcolumn}% Align table columns on decimal point
\usepackage{bm}% bold math
\usepackage[utf8]{inputenc}
\usepackage[T1]{fontenc}
\usepackage{mathptmx}
\usepackage{etoolbox}

\usepackage{xcolor}
\usepackage[
colorlinks=true,
linkcolor=blue,
citecolor=blue,
urlcolor=blue
]{hyperref}

\begin{document}
	
%\preprint{AIP/123-QED}

\title{Enantiomer-specific pumping of chiral molecules in high-$J$ rotational states}

\author{Fen Zou}
\affiliation{Center for Theoretical Physics \& School of Physics and Optoelectronic Engineering, Hainan University, Haikou 570228, China.}

\author{Yan Gao}
\affiliation{Center for Theoretical Physics \& School of Physics and Optoelectronic Engineering, Hainan University, Haikou 570228, China.}

\author{Peng Zhang}
\thanks{Electronic mail: pengzhang@ruc.edu.cn}
\affiliation{School of Physics, Renmin University of China, Beijing, 100872, China.}%
\affiliation{Key Laboratory of Quantum State Construction and Manipulation (Ministry of Education), Renmin University of China, Beijing, 100872, China}

\date{\today}% It is always \today, today,

\begin{abstract}

Enantiomer-specific state transfer (ESST) based on electric-dipole transitions has attracted considerable interest in the studies of manipulating molecules with electromagnetic fields. It also holds promise as a powerful tool for enantiomer detection and separation. To date, experimental demonstrations of ESST have focused mainly on molecules in the lowest rotational angular-momentum states, namely, $J=0$ and $J=1$. Realistic molecular samples, however, may contain appreciable populations in higher-$J$ rotational states, making it desirable to develop ESST schemes applicable to such states. Nevertheless, one cannot derive such schemes via a simple and straightforward extension of existing low-$J$ schemes, because higher-$J$ manifolds contain more magnetic sub-levels and involve more complicated networks of electric-dipole transitions. Here, we propose an ESST scheme for chiral molecules involving the $J=1$ and $J=2$ rotational states. Specifically, our scheme is based on enantiomer-specific pumping driven by three linearly polarized microwave fields and one laser beam. Under this scheme, one enantiomer is pumped into an enantiomer-specific dark state within a sub manifold with $J=1,2$, whereas the other enantiomer is dissipatively pumped out of this manifold. Our strategy can be  extended to systems involving rotational states with even higher angular momenta.

\end{abstract}

\maketitle

\section{Introduction}

A chiral molecule cannot be superimposed on its mirror image by any combination of translations and rotations, and therefore usually exists in two mirror-image configurations known as enantiomers. Although two enantiomers have nearly identical physical properties, their different spatial configurations can lead to pronounced differences in chemical reactivity and biological activity~\cite{Mezey1991}. Reliable detection, manipulation, and separation of molecular enantiomers are therefore of central importance in chemistry, biology, and pharmaceutical science. Conventional spectroscopic methods for chiral detection include circular dichroism~\cite{Berova2000}, vibrational circular dichroism~\cite{Stephens1985}, and Raman optical activity~\cite{YananHe2011}. These methods, however, usually rely on weak magnetic-dipole or electric-quadrupole interactions between molecules and light fields, which limits their sensitivity, especially for small molecular samples.

In recent years, electric-dipole-mediated approaches have attracted considerable attention for the detection and separation of molecular enantiomers. Enantiomer-specific state transfer (ESST) serves as a key step in a major class of these approaches~\cite{Li2008,Jia2010,Leibscher2019,Vitanov2019,Wu2019,Ye2019,Torosov2020,Torosov2020s,Wu2020,Guo2022,Leibscher2022,Liu2022,Cheng2023,Liu2024,Eibenberger2017,Lee2022,Lee2024a,Lee2024,Zou2025}. ESST relies on a cyclic scheme of electric-dipole transitions connecting three internal molecular states. The product of the three corresponding transition dipole moments has opposite signs for the two enantiomers~\cite{Kral2001,Kral2003}. This sign difference enables molecules of the two enantiomers, initially prepared in the same internal state, to be transferred into different final states. Once ESST is achieved, it can be further exploited for enantiomer detection~\cite{Jia2011,Patterson2013,Patterson2013a,Patterson2014,Shubert2014,Shubert2015,Lobsiger2015,Yachmenev2016,Ye2019a,Xu2020,Ye2021,Cai2022,Chen2022,Kang2023,Ye2023} and spatial separation~\cite{Li2007,Li2010,Eilam2013,Liu2021,Chen2024}.

The cyclic transitions required for ESST can be driven by electromagnetic fields, which are near resonant with transitions between molecular eigenstates. In realistic molecules, however, each energy level generally contains multiple degenerate states. Consequently, a given field may simultaneously drive several transitions rather than selectively coupling a single pair of states. This multilevel structure must therefore be taken into account when designing experimentally feasible ESST schemes.

To date, ESST has been experimentally demonstrated for molecules in the lowest rotational states, involving the $J=0$ and $J=1$ manifolds~\cite{Eibenberger2017,Lee2022,Lee2024a}. For these states, three appropriately polarized electromagnetic fields can selectively isolate three relevant molecular states and couple them through a closed cycle of electric-dipole transitions~\cite{Ye2018}. In realistic molecular samples, however, the temperature may not be sufficiently low for all molecules to occupy only these lowest rotational states, and a finite fraction may remain thermally populated in higher-$J$ states. It is therefore desirable to develop ESST schemes applicable to higher rotational manifolds. Such manifolds contain substantially more degenerate magnetic sublevels~\cite{Lee2024,Leibscher2022,Leibscher2022s}, giving rise to a more complex network of electric-dipole transitions. Designing ESST schemes for these manifolds thus requires the full multilevel transition structure to be considered, and cannot  be achieved through a straightforward extension of existing schemes for the $J=0$ and $J=1$ states.

\begin{figure}
 	\includegraphics[width=0.5\textwidth]{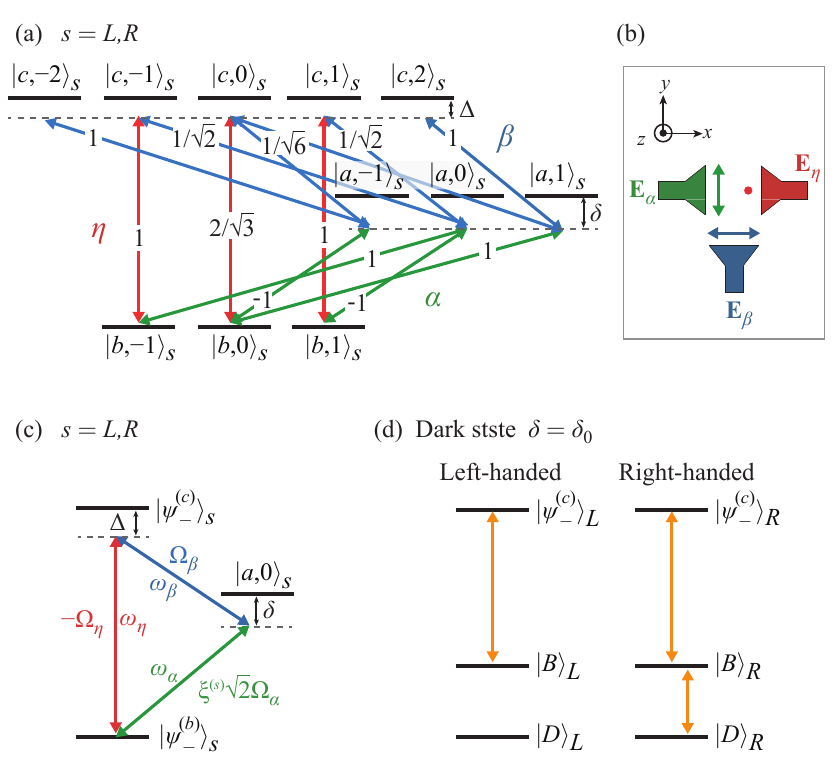}
 	\caption{\label{fig1}
 		{\bf (a):} Rotational states in the $a$-, $b$- and $c$-levels of molecule $s$ ($s=L,R$). The three MWs $\alpha$, $\beta$, and $\eta$ can induce transitions between these states. The numbers labeling lines of the same color indicate the relative Clebsch--Gordan coefficients of the corresponding transitions.
 		{\bf (b):} Polarizations of the three MWs.
 		{\bf (c):} Effective cyclic three-level configuration described by $H_{I1}^{(s)}$.
 		{\bf (d):} Enantiomer-dependent coupling configurations for the case with $\delta=\delta_{0}$, with $\delta_0$ being defined in Eq. (\ref{delta}). For the left-handed molecule, $|D\rangle_L$ is decoupled from both $|B\rangle_L$ and $|\psi_-^{(c)}\rangle_L$ and therefore forms a dark state. For the right-handed molecule, $|D\rangle_R$ is coupled to $|B\rangle_R$, which is in turn coupled to $|\psi_-^{(c)}\rangle_R$; therefore, $|D\rangle_R$ is not a dark state.}
 \end{figure}

In this work, we propose a scheme for realizing enantiomer-specific pumping (ESP), a form of ESST~\cite{Zou2024}, in chiral molecules involving rotational states with $J=1$ and $J=2$. Specifically, we consider two rotational levels with $J=1$ and one rotational level with $J=2$, which contain a total of eleven orthogonal magnetic states, as illustrated in Fig.~\ref{fig1}(a). The three rotational levels are coupled by three microwave fields $\alpha$, $\beta$ and $\eta$, which are linearly polarized along the $y$, $x$, and $z$ directions of the lab frame, respectively [Fig.~\ref{fig1}(b)]. By taking into account all the eleven magnetic states and the relevant near-resonant electric-dipole transitions, we show that the system contains a three-state subsystem with cyclic transitions being induced by the microwaves. Two of the three states forming this cycle are coherent superpositions of multiple magnetic states.
Furthermore, by combining this cyclic three-state subsystem with laser excitation and spontaneous decay, one can realize ESP~\cite{Zou2024}, whereby one enantiomer is pumped into an enantiomer-specific dark state, while the other is dissipatively pumped out of all three rotational levels.

Our scheme can serve as a basis for the detection and separation of enantiomers in chiral molecules occupying the $J=1$ and $J=2$ rotational states. It also demonstrates that ESST in higher-angular-momentum rotational manifolds can be achieved with an experimental setup of complexity comparable to that used for the $J=0$ and $J=1$ states, involving only three microwave fields and one laser beam. These results suggest that developing further ESST schemes of similar experimental complexity for chiral molecules in higher rotational states is a promising direction for future research.

The remainder of this paper is organized as follows. In Sec.~\ref{sec2}, we introduce the  setup, derive the effective cyclic three-level subsystem, and identify the chiral dark state. In Sec.~\ref{sec3}, we present the ESP scheme, analyze its robustness against collisional relaxation, and discuss its application to enantiodetection. We summarize our results in Sec.~\ref{sec4}. Some details of our calculations are given in the Appendix.

\section{Setup and Enantiomer-Specific Dark State}\label{sec2}

\subsection{Setup}
\label{sec2a}

As shown in Fig.~\ref{fig1}(a), we consider a chiral molecule with handedness $s$ ($s=L, R$) in three rotational levels $a$, $b$ and $c$ of the ground electronic and vibrational state. Specifically, each level includes several degenerate rotational states, as shown in the following:
\begin{eqnarray}
{\rm level-}a:\ \ |a,M\rangle_s&:=&|1, 0,M\rangle_s,\ \ \ \ \ \ (M=0,\pm 1);\\
{\rm level-}b:\ \ |b,M\rangle_s&:=&|1,-1,M\rangle_s,\ \ \ (M=0,\pm 1);\\
{\rm level-}c:\ \ |c,M\rangle_s&:=&|2,-2,M\rangle_s,\ \ \ (M=0,\pm 1,\pm 2).
\end{eqnarray}
Here we use the $|J,\tau,M\rangle_{s}$ notation~\cite{Ye2018}, with $J$ and $M$ 
being the quantum number of the angular momentum and its component along the $z$-axis of the lab frame, and 
$\tau$ labeling states within a given $J$ manifold in order of increasing energy. We further denote the self energies of the levels $a$, $b$, and $c$ as ($\hbar=1$) $\omega_{a,b,c}$, respectively. Clearly, we have [Fig.~\ref{fig1}(a)]
\begin{eqnarray}
\omega_c>\omega_a>\omega_b.
\end{eqnarray}

Additionally, three microwaves (MWs) $\alpha$, $\beta$, and $\eta$ are applied to induce electric-dipole couplings between three levels. Specifically, the polarization [Fig.~\ref{fig1}(b)] and induced coupling of these MWs are:
\begin{eqnarray}
{\rm MW}\ \ \ \ &{\rm polarization}&\ \ \ \ {\rm induced\ coupling}\nonumber\\
\alpha\ \ \ \ &{\bf e}_y&\ \ \ \ a\leftrightarrow b\ {\rm coupling}\nonumber\\
\beta\ \ \ \ &{\bf e}_x&\ \ \ \ a\leftrightarrow c\ {\rm coupling}\nonumber\\
\eta\ \ \ \ &{\bf e}_z&\ \ \ \ b\leftrightarrow c\ {\rm coupling}\nonumber
\end{eqnarray}
with
${\bf e}_{l}$ $(l=x,y,z)$ being the unit vector along the $l$ axis of the lab frame.
Accordingly, the electric field of the MW $\zeta$ ($\zeta=\alpha,\beta,\eta$) can be expressed as
\begin{eqnarray}
{\bm E}_{\zeta}={\mathcal{E}}_{\zeta}e^{-i(\omega_{\zeta}t+\varphi_{\zeta})}{\bf e}_{\zeta}+\mathrm{c.c.};\ \ 
(\zeta=\alpha,\beta,\eta),
\end{eqnarray}
where $\mathcal{E}_{\zeta}>0$, $\omega_{\zeta}$, and $\varphi_{\zeta}$ are the amplitude, angular frequency, and initial phase of the MW $\zeta$ $(\zeta=\alpha,\beta,\eta)$, respectively.
${\bf e}_{\alpha}={\bf e}_{y}$, ${\bf e}_{\beta}={\bf e}_{x}$, and ${\bf e}_{\eta}={\bf e}_{z}$ are the polarization vectors. Moreover, the frequencies $\omega_{\alpha,\beta,\eta}$ of the three beams satisfy
\begin{eqnarray}
\omega_\alpha+\omega_\beta=\omega_\eta.
\end{eqnarray}

\subsection{Enantiomer-specific dark state}

Now we consider the single-molecule Hamiltonian  of our system, which includes the self-energy of the rotational state and the MW-induced couplings. Taking into account the   polarization directions of the MWs, as well as the selection rule and the Clebsch–Gordan coefficients of these couplings [Fig.~\ref{fig1}(a)], we find that (Appendix~\ref{appa}) in the rotating frame the Hamiltonian of a molecule with handedness $s$ ($s=L,R$) is given by
\begin{eqnarray}\label{HamI}
	H_{I}^{(s)}&=&
	\Delta\sum_{M=-2}^{2}|c,M\rangle_{s}\langle c,M|+
	\delta\sum_{M=-1}^{1}|a,M\rangle_{s}\langle a,M|    \nonumber \\ 
	&&+\frac{1}{2}\Big [e^{i\phi_{s}}\Omega_{\alpha}(|a,1\rangle_{s}\langle b,0|-|a,-1\rangle_{s}\langle b,0|) \nonumber\\
	&&+e^{i\phi_{s}}\Omega_{\alpha}(|a,0\rangle_{s}\langle b,-1|-|a,0\rangle_{s}\langle b,1|) \nonumber\\
	&&+\Omega_{\beta}(|c,2\rangle_{s}\langle a,1|-|c,-2\rangle_{s}\langle a,-1|) \nonumber\\
	&&+\frac{\Omega_{\beta}}{\sqrt{2}}(|c,1\rangle_{s}\langle a,0|-|c,-1\rangle_{s}\langle a,0|) \nonumber\\
	&&+\frac{\Omega_{\beta}}{\sqrt{6}}(|c,0\rangle_{s}\langle a,-1|-|c,0\rangle_{s}\langle a,1|)  \nonumber\\	
	&&+\Omega_{\eta}(|c,1\rangle_{s}\langle b,1|+|c,-1\rangle_{s}\langle b,-1|) \nonumber\\ 
	&&+\frac{2\Omega_{\eta}}{\sqrt{3}}|c,0\rangle_{s}\langle b,0|+\mathrm{H.c.}\Big],  \quad (s=L,R),\label{hi}
\end{eqnarray}
where 
\begin{eqnarray}
\Delta&=&\omega_{c}-\omega_{b}-\omega_{\eta};\\
\delta&=&\omega_{a}-\omega_{b}-\omega_{\alpha},
\end{eqnarray}
are the one- and two-photon detuning, respectively. Here $\Omega_{\alpha,\beta,\eta}>0$ are the norm of the reduced matrix element of the transition induced by the MWs $\alpha$,  $\beta$, and $\eta$, respectively. Additionally, the overall phase $\phi_s$ $(s=L,R)$ is controlled by the phase difference $\phi_\alpha+\phi_\beta-\phi_\eta$ of the three MWs, and satisfies $\phi_{R}=\phi_{L}+\pi$, as shown in Appendix~\ref{appa}. Without loss of generality, here we assume
\begin{eqnarray}
\phi_L=0;\ \ \phi_R=\pi.
\end{eqnarray}

For the convenience of the following discussions, we introduce the states $|\psi_{\pm}^{(b,c)}\rangle_{s}$ $(s=L,R)$  as
\begin{eqnarray}\label{SepStates}
|\psi_{\pm}^{(b)}\rangle_{s}&=&(|b,-1\rangle_{s}\pm|b,1\rangle_{s})/\sqrt{2};\nonumber\\
|\psi_{\pm}^{(c)}\rangle_{s}&=&(|c,1\rangle_{s}\pm|c,-1\rangle_{s})/\sqrt{2},\quad (s=L,R).
\end{eqnarray}
Moreover, we can decompose the eleven dimensional Hilbert ${\mathbb H}^{(s)}$ of molecular $s$ ($s=L,R$) as the direct sum of two subspaces, i.e., 
\begin{eqnarray}
{\mathbb H}^{(s)}={\mathbb H}^{(s)}_1\oplus{\mathbb H}^{(s)}_2,
\end{eqnarray}
where ${\mathbb H}^{(s)}_1$ and ${\mathbb H}^{(s)}_2$ are three- and eight-dimensional subspaces, respectively. Specifically,
\begin{eqnarray}
{\mathbb H}^{(s)}_1\!\!&=&\!\!{\rm span}\{|a,0\rangle_{s}, |\psi_{-}^{(b)}\rangle_{s}, |\psi_{-}^{(c)}\rangle_{s}\},\\[4pt]
{\mathbb H}^{(s)}_2\!\!&=&\!\!{\rm span}\{|a,\pm 1\rangle_{s}, |\psi_{+}^{(b)}\rangle_{s}, |\psi_{+}^{(c)}\rangle_{s},|b,0\rangle_{s},|c,0\rangle_s,|c,\pm 2\rangle_{s}\}.\nonumber\\
\end{eqnarray}
A straightforward calculation shows that the Hamiltonian $H_I^{(s)}$ in Eq.~(\ref{hi}) does not couple $\mathbb{H}_1^{(s)}$ to $\mathbb{H}_2^{(s)}$, as illustrated in Fig.~\ref{fig1}(c). In other word, we have
\begin{eqnarray}
H_{I}^{(s)}={\rm P}_1H_{I}^{(s)}{\rm P}_1+{\rm P}_2H_{I}^{(s)}{\rm P}_2,
\end{eqnarray}
where ${\rm P}_1$ and ${\rm P}_2$ are the projection operators of the subspaces $\mathbb{H}_1^{(s)}$ and $\mathbb{H}_2^{(s)}$, respectively. 

Now we focus on the  Hamiltonian in the space $\mathbb{H}_1^{(s)}$, which is given by
\begin{eqnarray}
	H_{I1}^{(s)}&:=&{\rm P}_1H_{I}^{(s)}{\rm P}_1\nonumber\\
	&\ \ =&
	\Delta|\psi_{-}^{(c)}\rangle_{s}\langle\psi_{-}^{(c)}|+\delta|a,0\rangle_{s}\langle a,0|\nonumber\\
	&&+\frac{1}{2}\Big(\xi^{(s)}\sqrt{2}\Omega_{\alpha}|a,0\rangle_{s}\langle \psi_{-}^{(b)}| +\Omega_{\beta}|\psi_{-}^{(c)}\rangle_{s}\langle a,0|\nonumber\\
	&&-\Omega_{\eta}|\psi_{-}^{(c)}\rangle_{s}\langle \psi_{-}^{(b)}|+\mathrm{H.c.}\Big) ,       \quad(s=L,R) ,  \label{hi1}
\end{eqnarray}
where $\xi^{(L)}=1$ and $\xi^{(R)}=-1$.
Equation~(\ref{hi1}) yields that the MWs can induce closed cyclic transitions between the three states $\{|a,0\rangle_{s}, |\psi_{-}^{(b)}\rangle_{s}, |\psi_{-}^{(c)}\rangle_{s}\}$, as shown in Fig.~\ref{fig1}(c).
Moreover, we define two orthogonal states in the subspace $\mathbb{H}_{1}^{(s)}$ as
\begin{eqnarray}
	|D\rangle_{s}
	&=&
	\mathcal{Z}^{-1}
	\left[
	\Omega_{\eta}|a,0\rangle_{s}
	+\Omega_{\beta}|\psi_{-}^{(b)}\rangle_{s}
	\right],
	\nonumber\\
	|B\rangle_{s}
	&=&
	\mathcal{Z}^{-1}
	\left[
	\Omega_{\beta}|a,0\rangle_{s}
	-\Omega_{\eta}|\psi_{-}^{(b)}\rangle_{s}
	\right],
\end{eqnarray}
where $\mathcal{Z}=\sqrt{\Omega_{\beta}^{2}+\Omega_{\eta}^{2}}$.
The three states $\{|\psi_{-}^{(c)}\rangle_{s},|D\rangle_{s},|B\rangle_{s}\}$ thus form a new orthogonal basis of $\mathbb{H}_{1}^{(s)}$. In this basis, the Hamiltonian $H_{I1}^{(s)}$ can be written as
\begin{eqnarray}
	H_{I1}^{(s)}
	&=&
	\Delta|\psi_{-}^{(c)}\rangle_{s}\langle\psi_{-}^{(c)}|
	+\varepsilon_{D}^{(s)}|D\rangle_{s}\langle D|
	+\varepsilon_{B}^{(s)}|B\rangle_{s}\langle B|
	\nonumber\\
	&&+
	\left[
	\frac{\mathcal{Z}}{2}|\psi_{-}^{(c)}\rangle_{s}\langle B|
	+\Lambda_{s}|B\rangle_{s}\langle D|
	+\mathrm{H.c.}
	\right],\label{HI1s}
\end{eqnarray}
where
\begin{eqnarray}
	\varepsilon_{D}^{(s)}
	&=&
	\frac{\delta\Omega_{\eta}^{2}+\sqrt{2}\xi^{(s)}\Omega_{\alpha}\Omega_{\eta}\Omega_{\beta}}{\mathcal{Z}^{2}},\\
	\varepsilon_{B}^{(s)}
	&=&
	\frac{\delta\Omega_{\beta}^{2}-\sqrt{2}\xi^{(s)}\Omega_{\alpha}\Omega_{\eta}\Omega_{\beta}}{\mathcal{Z}^{2}},
	\end{eqnarray}
and	
\begin{eqnarray}
	\Lambda_{s}
	&=&{}_{s}\langle B|H_{I1}^{(R)}|D\rangle_{s}=
	\frac{1}{\mathcal{Z}^{2}}
	\left[
	\delta\Omega_{\eta}\Omega_{\beta}
	+\frac{\xi^{(s)}}{\sqrt{2}}\Omega_{\alpha}
	\left(\Omega_{\beta}^{2}-\Omega_{\eta}^{2}\right)
	\right].\nonumber\\
\end{eqnarray}
Notice that Eq. (\ref{HI1s}) yields that
\begin{equation}
	{}_{s}\langle\psi_{-}^{(c)}|H_{I1}^{(s)}|D\rangle_{s}=0,\label{hi10}
	\end{equation}
i.e., there is no direct coupling between $|D\rangle_{s}$ and $|\psi_{-}^{(c)}\rangle_{s}$.

\begin{figure*}
	\includegraphics[width=0.75\textwidth]{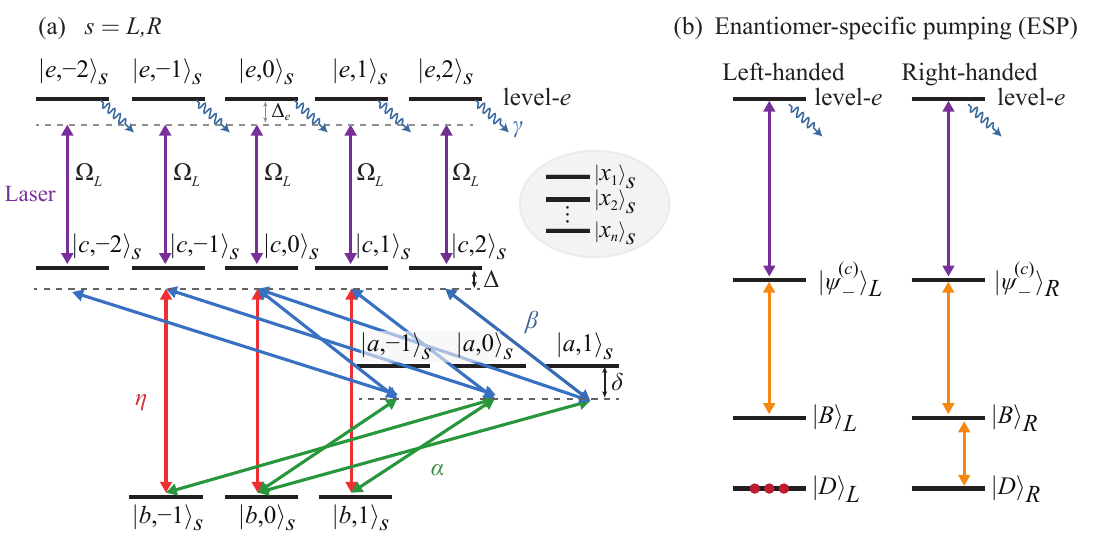}
	\caption{\label{fig2}
	{\bf (a):} Rotational levels and transitions involved in  the ESP scheme.  {\bf (b):} Schematic  illustration of the underlying mechanism of ESP. Further details are provided in the main text.
	}
\end{figure*}
We now choose the detuning as
\begin{equation}\label{delta}
	\delta
	=
	\frac{\Omega_{\alpha}\left(\Omega_{\eta}^{2}-\Omega_{\beta}^{2}\right)}{\sqrt{2}\Omega_{\eta}\Omega_{\beta}}:=\delta_{0}\neq0.
\end{equation}
For the left-handed molecule, for which $\xi^{(L)}=1$, this condition, together with Eq. (\ref{hi10}),  gives
\begin{equation}
	{}_{L}\langle\psi_{-}^{(c)}|H_{I1}^{(L)}|D\rangle_{L}
	=
	{}_{L}\langle B|H_{I1}^{(L)}|D\rangle_{L}
	=0.
\end{equation}
Therefore, $|D\rangle_{L}$ is decoupled from the other two basis states and is an eigenstate of $H_{I1}^{(L)}$ [Fig.~\ref{fig1}(d), left]. Because this eigenstate contains no component of the $c$ level, it is a dark state of the left-handed molecule.

For the right-handed molecule, however, $\xi^{(R)}=-1$, and the same detuning condition yields
\begin{equation}
	{}_{R}\langle B|H_{I1}^{(R)}|D\rangle_{R}\neq0.
\end{equation}
Thus, $|D\rangle_{R}$ is coupled to $|B\rangle_{R}$, which is in turn coupled to $|\psi_{-}^{(c)}\rangle_{R}$  [Fig.~\ref{fig1}(d), right]. It is straightforward to show that, under this condition, all three eigenstates of $H_{I1}^{(R)}$ contain a component of the $c$ level. Consequently, only the left-handed molecule possesses the dark state $|D\rangle_{L}$, whereas the right-handed molecule has no dark state. We therefore refer to $|D\rangle_{L}$ as a chiral dark state. Similarly, when $\delta=-\delta_{0}$, $|D\rangle_{R}$ becomes a dark state of the right-handed molecule, whereas the left-handed molecule has no corresponding dark state.

In our previous work, we discussed chirality-dependent dark states within a generic cyclic three-level model~\cite{Zou2024}. The above analysis demonstrates that such a chiral dark state can be realized in realistic chiral molecules occupying high-$J$ rotational states.

\section{ESP for $J=1,2$ molecules}\label{sec3}

\subsection{Scheme}

Building on the chiral dark state identified above, we can implement the ESP scheme, whose underlying principle was proposed in our previous work on a three-level system with cyclic electric-dipole transitions~\cite{Zou2024}. Specifically, 
in addition to the three MWs introduced above, we introduce a $\pi$-polarized laser to couple the states $|c,M\rangle_s$ in the $c$ level to the states $|e,M\rangle_s$ in an electronically excited level, where $M=0,\pm1,\pm2$ [Fig.~\ref{fig2}(a)]. Molecules excited to the $e$ level can subsequently decay through spontaneous emission to rovibrational levels in the electronic ground state, including the $a$, $b$, and $c$ levels, as well as other levels collectively denoted by $|x_1\rangle_s,\ldots,|x_n\rangle_s$ [Fig.~\ref{fig2}(a)]. 

In the following, we take the case with $\phi_L=0$ and $\delta=\delta_0$ as an example to illustrate this scheme. In this case, for the right-handed molecules ($s=R$), all rotational states in the $a$ and $b$ levels are coupled to the $c$ level by the MWs. Therefore, under the combined action of the MWs and laser, molecules initially occupying any state in the $a$, $b$, or $c$ level can be transferred through the $c$ level to the electronically excited $e$ level, and then decay spontaneously to rovibrational levels in the electronic ground state, including not only the levels $a$, $b$ and $c$,  but also the levels $|x_1\rangle_R,\ldots,|x_n\rangle_R$. Repetition of this excitation--decay cycle eventually pumps all right-handed molecules out of the $a$, $b$, and $c$ levels and into other rovibrational levels of the electronic ground state [Fig.~\ref{fig2}(b), right]. Consequently, no right-handed molecules remain in the $a$, $b$, or $c$ level.

For the left-handed molecules ($s=L$), by contrast, the MWs cannot couple the dark state $|D\rangle_L$ to any other state. Molecules occupying $|D\rangle_L$ therefore cannot be transferred to the $c$ level by the MWs and, consequently, cannot be excited to the $e$ level by the laser or undergo the subsequent spontaneous-emission process. Thus, these molecules remain within the $a$ and $b$ levels [Fig.~\ref{fig2}(b), left].
Additionally, molecules in all other states in the $a$, $b$, and $c$ levels are pumped out through the same excitation--decay cycle as for the right-handed molecule. 

In summary, for $\phi_L=0$ and $\delta=\delta_0$, the combined action of the MWs and laser pumps all right-handed molecules out of the $a$- and $b$-levels, whereas a fraction of the left-handed molecules (i.e., those in the chiral dark state $|D\rangle_L$)
remains trapped in these two levels. Thus, ESP is realized.

In a realistic molecular gas, intermolecular collisions are unavoidable and can induce both population transfer among internal states and decoherence. In particular, inelastic collisions may transfer molecules from the $|x_{1}\rangle_{s},\ldots,|x_{n}\rangle_{s}$ levels back to the $a$-, $b$-, and $c$-levels. As a result, the populations of the right-handed molecules in the $a$- and $b$-levels remain finite even in the steady state. Nevertheless, provided that the driving strengths of the MWs and laser, as well as the spontaneous decay rates of the states $|e,M\rangle_{L(R)}$ ($M=0,\pm1,\pm2$), are much larger than the collisional relaxation rate, the populations of the left-handed molecules in the $a$- and $b$-levels remain significantly higher than those of the right-handed molecules in the corresponding states. Therefore, efficient ESP remains achievable even in the presence of collisions.

\begin{figure}
	\includegraphics[width=0.45\textwidth]{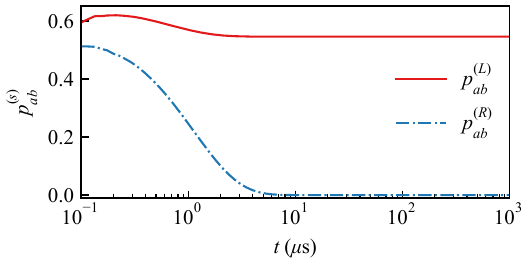}
	\caption{\label{fig3}
		Time evolution of the total populations $p_{ab}^{(L,R)}$ in the $a$- and $b$-levels. Here we show the result for the case with $\phi_L=0$, $\delta=\delta_0$, $\Delta=\Delta_e=0$, $\Omega_{\alpha}/2\pi=\Omega_{\beta}/2\pi=10\,\mathrm{MHz}$, $\Omega_{\eta}/2\pi=6\,\mathrm{MHz}$, $\Omega_L/2\pi=20\,\mathrm{MHz}$, $\gamma/2\pi=10\,\mathrm{MHz}$, and $\kappa=0$. Moreover, the number of $x$ states is $n=1$.
		The initial state is an equal incoherent mixture of the eleven rotational states in the $a$, $b$, and $c$ levels, i.e.,
		$\rho^{(s)}(0)=\big[\sum_{\mu=a,b}\sum_{M=-1}^{1}|\mu,M\rangle_s\langle\mu,M|+\sum_{M=-2}^{2}|c,M\rangle_s\langle c,M|\big]/11$.
}
\end{figure}

\subsection{Illustration}

\begin{figure}
	\includegraphics[width=0.48\textwidth]{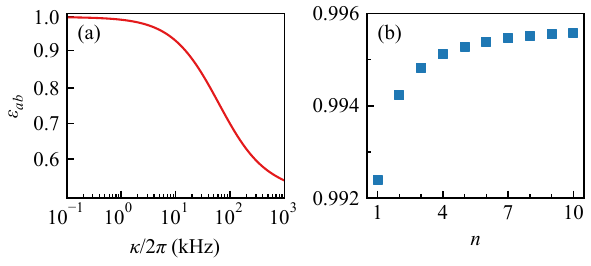}
	\caption{\label{fig4} Steady-state left-handed enantiomeric purity $\varepsilon_{ab}$. 
	{\bf (a):} $\varepsilon_{ab}$ as a function of the collisional relaxation rate $\kappa$, with $n=1$.
	{\bf (b):}  $\varepsilon_{ab}$ as a function of the number $n$ of $x$ states, with $\kappa/2\pi=1\,\mathrm{kHz}$. Other parameters are the same as those in Fig.~\ref{fig3}.}
\end{figure}

To further demonstrate our ESP scheme, we figure out the evolution of the single-molecule density matrix $\rho^{(s)}$  ($s=L,R$) by
 numerically solving the quantum master equation  for our current example with $\phi_L=0$ and $\delta=\delta_0$. This master equation can be expressed as
\begin{equation}
	\frac{d\rho^{(s)}}{dt}=-i[H^{\prime(s)}_{I},\rho^{(s)}]+\mathcal{L}[\rho^{(s)}]+\mathcal{D}[\rho^{(s)}].\label{ME}
\end{equation}
Here, the Hamiltonian $H^{\prime(s)}_{I}$ is given by
\begin{eqnarray}
	H^{\prime(s)}_{I} &=& H^{(s)}_{I}+(\Delta_{e}+\Delta)\sum_{M=-2}^{2}|e,M\rangle_{s}\langle e,M|\nonumber\\
	&&+\frac{\Omega_{L}}{2}\sum_{M=-2}^{2}(|c,M\rangle_{s}\langle e,M|+\mathrm{H.c.}), 
\end{eqnarray} 
where $H^{(s)}_{I}$ is given by Eq.~(\ref{HamI}), and $\Omega_{L}$ denotes the Rabi frequency of the laser-driven transition $|c,M\rangle_{s}\leftrightarrow|e,M\rangle_{s}$, and $\Delta_{e}=\omega_{e}-\omega_{c}-\omega_{L}$ is the laser detuning, with $\omega_{e}$ being energy of the  $e$ level and $\omega_{L}$ being the angular frequency of the laser. For simplicity, we assume that all laser-induced $c\leftrightarrow e$ transitions share the same Rabi frequency $\Omega_{L}$. 
Furthermore, in Eq. (\ref{ME}) the superoperators $\mathcal{L}[\rho^{(s)}]$ and $\mathcal{D}[\rho^{(s)}]$ describe spontaneous decay from $|e,M\rangle_{s}$ and intermolecule collisions, respectively. In general, both processes involve multiple channels with different final states, and their branching ratios depend on the molecular structure. For simplicity, we assume equal branching ratios for all spontaneous-decay and collisional channels. Under this assumption, the superoperator $\mathcal{L}[\rho^{(s)}]$ is given by~\cite{Scully1997}
\begin{eqnarray}
	\mathcal{L}[\rho^{(s)}]&=&\frac{\gamma}{(n+11)}\Big[\sum_{M=-1}^{1}\sum_{M^{\prime}=-2}^{2}(\mathcal{L}_{|b,M\rangle_{s}\langle e,M^{\prime}|}[\rho^{(s)}]\nonumber\\ &&+\mathcal{L}_{|a,M\rangle_{s}\langle e,M^{\prime}|}[\rho^{(s)}] )+\sum_{M,M^{\prime}=-2}^{2}\mathcal{L}_{|c,M\rangle_{s}\langle e,M^{\prime}|}[\rho^{(s)}]\nonumber\\
	&&+\sum_{k=1}^{n}\sum_{M^{\prime}=-2}^{2}\mathcal{L}_{|x_{k}\rangle_{s}\langle e,M^{\prime}|}[\rho^{(s)}] \Big], \quad \nonumber\\
	&&\hspace{2.7cm}(s=L,R),
\end{eqnarray}
where $\gamma$ is the spontaneous decay rate of the excited state $|e,M\rangle_{s}$ and $\mathcal{L}_{o}[\rho^{(s)}]\equiv[2o\rho^{(s)}o^{\dagger}-o^{\dagger}o\rho^{(s)}-\rho^{(s)}o^{\dagger}o]/2$. Note that here we neglect the spontaneous transitions among $|b,M\rangle_{s}$, $|a,M\rangle_{s}$, $|c,M\rangle_{s}$, and $|x_{1}\rangle_{s},\ldots,|x_{n}\rangle_{s}$ for $s=L,R$. Moreover, the collisional relaxation is modeled by~\cite{Zou2024}
\begin{equation}
	\mathcal{D}[\rho^{(s)}]=\kappa\Big[\frac{\hat{I}}{n+16}-\rho^{(s)} \Big],
\end{equation}
where $\hat{I}$ is the identity operator and $\kappa$ denotes the collision relaxation rate.

By solving Eq. (\ref{ME}), we can derive the 
time evolution of the total population in the $a$- and $b$-levels,  i.e., 
\begin{eqnarray}
p_{ab}^{(s)}(t)=\sum_{\mu=a,b}\sum_{M=-1}^{1}{}_{s}\langle\mu,M|\rho^{(s)}(t)|\mu,M\rangle_{s},\ \ (s=L,R).
\end{eqnarray}
Figure~\ref{fig3} shows $p_{ab}^{(L,R)}(t)$ for a typical case with $\phi_L=0$, $\delta=\delta_0$, $n=1$,
and negligible collision relaxation rate ($\kappa=0$). It was clearly shown that, as expected in our above analysis, the probability  $p_{ab}^{(R)}$ for the right-handed molecule decreases to  zero in the long-time limit, whereas $p_{ab}^{(L)}$ approaches a finite steady-state value. 
As mentioned above, this difference originates from the chiral dark state $|D\rangle_L$ of the left-handed molecule, which is decoupled from all other states under these conditions. Note that the  population in the long-time limit is determined by the steady state with respect to the master equation (\ref{ME}), which is independent of the molecular initial state. As a result, the ESP scheme requires neither initial-state preparation nor precise timing control of the MWs and laser.

We further examine the effect induced by the collisional relaxation. To this end we introduce the steady-state left-handed enantiomeric purity $\varepsilon_{ab}$ as
\begin{eqnarray}
\varepsilon_{ab}=\frac{P_{ab}^{(L)}}{P_{ab}^{(L)}+P_{ab}^{(R)}},
\end{eqnarray}
where
$P_{ab}^{(s)}=\sum_{\mu=a,b}\sum_{M=-1}^{1}{}_{s}\langle\mu,M|\rho_{\mathrm{ss}}^{(s)}|\mu,M\rangle_{s}$
denotes the steady-state total population in the $a$- and $b$-levels for molecule $s$ ($s=L,R$), with 
$\rho_{\mathrm{ss}}^{(s)}=\rho^{(s)}(t\rightarrow\infty)$ being the corresponding steady-state density matrix.  
When the collisional relaxation is negligible (i.e., $\kappa= 0$), we have $P_{ab}^{(R)}=0$ and thus $\varepsilon_{ab}=1$. In contrast, 
when the collisional relaxation is non-negligible (i.e., $\kappa\neq 0$), according to the analysis in the above subsection, we have $P_{ab}^{(R)}\neq0$ and thus $\varepsilon_{ab}<1$. Therefore, this purity can qualitatively describe the influence of the collision relaxation on the ESP. In particular, this effect is small when $\varepsilon_{ab}\approx 1$ (i.e., $1-\varepsilon_{ab}\ll 1$), because the steady-state populations of the $a$- and $b$-levels still differ substantially between the two enantiomers under this condition.

In Fig.~\ref{fig4}(a) we show the purity $\varepsilon_{ab}$ for cases with non-zero $\kappa$, and other parameters being same as those of Fig.~\ref{fig3}. It is shown that, as expected in the above analysis, $\varepsilon_{ab}$ decreases from 1 when the collision relaxation rate $\kappa$ is increased from zero. Moreover, we have $\varepsilon_{ab}\approx 1$ when 
$\kappa$ is much less than the Rabi frequencies $\Omega_{\alpha,\beta,\eta,L}$ of the MWs and the laser, as well as the spontaneous emission rate $\gamma$ of the excited level. Specifically, for our example with $\Omega_{\alpha,\beta,\eta,L}/(2\pi)$ and $\gamma/(2\pi)$ being $6$-$20$ MHz, we have $\varepsilon_{ab}\gtrsim 0.99$ for $\kappa/2\pi\lesssim1\,\mathrm{kHz}$, and $\varepsilon_{ab}\gtrsim 0.9$ for $\kappa/2\pi\lesssim10\,\mathrm{kHz}$. Thus, high enantiomeric purity can be maintained provided that collisional relaxation is sufficiently weak, which is consistent with our analysis in the above subsection. This weak-collision condition may be approached experimentally by reducing the molecular density and temperature.

Furthermore, the results in Fig.~\ref{fig4}(a) were obtained for a single $x$-state ($n=1$). To assess the dependence on the number $n$ of $x$-states, in Fig.~\ref{fig4}(b) we show the purity $\varepsilon_{ab}$ as a function of $n$. As in the example studied in our previous work~\cite{Zou2024}, the purity $\varepsilon_{ab}$ increases only slightly with $n$ and gradually approaches saturation. The ESP scheme is therefore largely insensitive to the precise number of $x$-states.

\subsection{Applications of ESP}

As shown in our previous work~\cite{Zou2024}, the ESP scheme can be used for both enantiomer detection and the spatial separation of chiral molecules. Here, we again consider the system with $\phi_L=0$ as an example. As discussed above, when the MW detuning $\delta$ is set to $\delta_0$, only left-handed molecules remain in levels $a$ and $b$ after the ESP process. Thus, a laser that selectively ionizes or dissociates molecules in either level $a$ or level $b$ acts exclusively on left-handed molecules, enabling the spatial separation of left- and right-handed molecules.

\begin{figure}
	\includegraphics[width=0.45\textwidth]{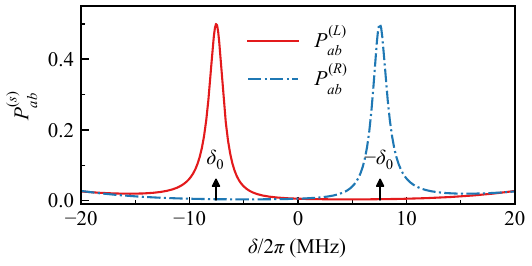}
	\caption{\label{fig5}
		Steady-state total populations $P_{ab}^{(s)}$ in the $a$- and $b$-levels for the left- and right-handed molecules ($s=L,R$, respectively) as a function of the detuning $\delta$. The arrows indicate $\delta=\pm\delta_0$. Here,  $\kappa/2\pi=1\,\mathrm{kHz}$, and other parameters are the same as those in Fig.~\ref{fig3}.}
\end{figure}

To measure the enantiomeric excess of a mixture of left- and right-handed molecules, one can prepare two samples of the mixture, labeled 1 and 2, containing equal total numbers of molecules. ESP is then performed on samples 1 and 2 with $\delta=\delta_0$ and $\delta=-\delta_0$, respectively, and the total number of molecules remaining in levels $a$ and $b$ is measured for each sample. The ratio of these measured populations in sample 1 to sample 2 equals the ratio of left- to right-handed molecules in the original mixture, from which the enantiomeric excess can be determined.

Applying the above approaches requires knowledge of $\delta_0$, which is determined by the molecular structure. However, as shown in our previous work~\cite{Zou2024}, ESP can still be used to measure the enantiomeric excess without the knowledge of the exact value of $\delta_0$. Specifically, for a mixture of the two enantiomers, one can scan the detuning $\delta$ by varying the MW frequencies and measure the steady-state total population in levels $a$ and $b$, $P_{ab}$, after ESP at each detuning. The resulting curve, $P_{ab}(\delta)$, exhibits two  peaks centered at $\delta=\pm\delta_0$, whose relative heights reflect the relative abundances of the two enantiomers. This approach therefore does not require prior knowledge of the exact peak positions. To illustrate this approach, in Fig.~\ref{fig5} we show the steady-state total populations $P_{ab}^{(L,R)}(\delta)$ for pure samples of left- and right-handed molecules for a typical case. Two distinct peaks, corresponding to the two enantiomers, are clearly visible.

\section{Summary and Discussions}\label{sec4}

In this work, we develop an ESP scheme for chiral molecules in the $J=1$ and $J=2$ rotational manifolds. As demonstrated in the preceding section, this scheme enables the detection and spatial separation of molecular enantiomers in these rotational states.

Our scheme has the same complexity as those for molecules in the $J=0$ and $J=1$ rotational states, requiring only three MW fields. This is because, as we found in Sec.~\ref{sec2}, the symmetry of the Clebsch--Gordan coefficients allows three specific states to be decoupled from the remaining magnetic sublevels under the applied MW fields, forming closed cyclic transitions, as shown in Fig.~\ref{fig1}(c). Two of these three states are superpositions of bare magnetic rotational states. This strategy for identifying closed three-state transition cycles can also be applied directly to molecules in higher rotational states. Our results thus suggest that ESP and other types of ESST may be realized using only three MW fields for molecules in high-$J$ manifolds, facilitating the manipulation of chiral molecules at relatively high temperatures.

\appendix

\section{Derivation of the Hamiltonian (\ref{hi})}\label{appa}

In this appendix, we derive the Hamiltonian $H_{I}^{(s)}$ ($s=L,R$) of Eq.~(\ref{hi}) in the rotating frame. 
We first notice that, in the Schr\"odinger picture, 
the Hamiltonian for the molecule interacting with the MWs, which is introduced in Sec.~\ref{sec2a} for detail, can be expressed as ($\hbar=1$)
\begin{eqnarray}\label{Ham2a}
	H^{(s)}_{\mathrm{Schr}}&=&\sum_{\mu=a,b}\sum_{M=-1}^{1}\omega_{\mu}|\mu,M\rangle_{s}\langle \mu,M|+\sum_{M=-2}^{2}\omega_{c}|c,M\rangle_{s}\langle c,M|    \nonumber \\ 
	&&+\frac{1}{2}\Big [\Omega_{\alpha}^{(s)}e^{-i\omega_{\alpha}t}(|a,1\rangle_{s}\langle b,0|-|a,-1\rangle_{s}\langle b,0|) \nonumber\\
	&&+\Omega_{\alpha}^{(s)}e^{-i\omega_{\alpha}t}(|a,0\rangle_{s}\langle b,-1|-|a,0\rangle_{s}\langle b,1|) \nonumber\\
	&&+\Omega_{\beta}^{(s)}e^{-i\omega_{\beta}t}(|c,2\rangle_{s}\langle a,1|-|c,-2\rangle_{s}\langle a,-1|) \nonumber\\
	&&+\frac{\Omega_{\beta}^{(s)}}{\sqrt{2}}e^{-i\omega_{\beta}t}(|c,1\rangle_{s}\langle a,0|-|c,-1\rangle_{s}\langle a,0|) \nonumber\\
	&&+\frac{\Omega_{\beta}^{(s)}}{\sqrt{6}}e^{-i\omega_{\beta}t}(|c,0\rangle_{s}\langle a,-1|-|c,0\rangle_{s}\langle a,1|)  \nonumber\\	
	&&+\Omega_{\eta}^{(s)}e^{-i\omega_{\eta}t}(|c,1\rangle_{s}\langle b,1|+|c,-1\rangle_{s}\langle b,-1|) \nonumber\\ 
	&&+\frac{2\Omega_{\eta}^{(s)}}{\sqrt{3}}e^{-i\omega_{\eta}t}|c,0\rangle_{s}\langle b,0|+\mathrm{H.c.}\Big].
\end{eqnarray}
Here the energies $\omega_{\alpha}$ and the states $|\alpha,M\rangle$ ($\alpha=a,b,c$) are defined in Sec.~\ref{sec2}, and 
the complex Rabi frequencies are defined as 
\begin{equation}
	\Omega_{\xi}^{(s)}=\Omega_{\xi}e^{i\phi_{\xi}^{(s)}},
	\qquad \Omega_{\xi}>0,
	\qquad \xi=\alpha,\beta,\eta.
\end{equation}
Here the phases $\phi_{\xi}^{(L,R)}$ ($\xi=\alpha,\beta,\eta$) are determined by the phases of the MWs, and satisfy 
\begin{equation}
	\phi_{\xi}^{(R)}=\phi_{\xi}^{(L)}+\pi.
\end{equation}

We further make the transformation
\begin{eqnarray}
|{a,M}\rangle_{s}&\rightarrow&e^{i\phi_{\beta}^{(s)}}|a,M\rangle_{s},  \quad M=0,\pm1,   \nonumber\\
|{b,M}\rangle_{s}&\rightarrow&e^{i\phi_{\eta}^{(s)}}|b,M\rangle_{s}, \quad M=0,\pm1.
\end{eqnarray}
After this transformation,  the Hamiltonian in Eq.~(\ref{Ham2a}) can be re-written as
\begin{eqnarray}
	H^{(s)}_{\mathrm{Schr}}&=&\sum_{\mu=a,b}\sum_{M=-1}^{1}\omega_{\mu}|{\mu,M}\rangle_{s}\langle {\mu,M}|+\sum_{M=-2}^{2}\omega_{c}|{c,M}\rangle_{s}\langle {c,M}|    \nonumber \\ 
	&&+\frac{1}{2}\Big [\Omega_{\alpha}e^{i\phi_{s}}e^{-i\omega_{\alpha}t}(|{a,1}\rangle_{s}\langle {b,0}|-|{a,-1}\rangle_{s}\langle {b,0}|) \nonumber\\
	&&+\Omega_{\alpha}e^{i\phi_{s}}e^{-i\omega_{\alpha}t}(|{a,0}\rangle_{s}\langle {b,-1}|-|{a,0}\rangle_{s}\langle {b,1}|) \nonumber\\
	&&+\Omega_{\beta}e^{-i\omega_{\beta}t}(|{c,2}\rangle_{s}\langle {a,1}|-|{c,-2}\rangle_{s}\langle {a,-1}|) \nonumber\\
	&&+\frac{\Omega_{\beta}}{\sqrt{2}}e^{-i\omega_{\beta}t}(|{c,1}\rangle_{s}\langle {a,0}|-|{c,-1}\rangle_{s}\langle {a,0}|) \nonumber\\
	&&+\frac{\Omega_{\beta}}{\sqrt{6}}e^{-i\omega_{\beta}t}(|{c,0}\rangle_{s}\langle {a,-1}|-|{c,0}\rangle_{s}\langle {a,1}|)  \nonumber\\	
	&&+\Omega_{\eta}e^{-i\omega_{\eta}t}(|{c,1}\rangle_{s}\langle {b,1}|+|{c,-1}\rangle_{s}\langle {b,-1}|) \nonumber\\ 
	&&+\frac{2\Omega_{\eta}}{\sqrt{3}}e^{-i\omega_{\eta}t}|{c,0}\rangle_{s}\langle {b,0}|+\mathrm{H.c.}\Big],
\end{eqnarray}
where the overall phase $\phi_{s}$ ($s=L,R$) is defined as
\begin{equation}
	\phi_{s}=\phi_{\alpha}^{(s)}+\phi_{\beta}^{(s)}-\phi_{\eta}^{(s)},\ \ \ \ \ (s=L,R),
\end{equation}
and satisfy
\begin{equation}
	\phi_{R}=\phi_{L}+\pi
	\quad (\mathrm{mod}\ 2\pi).
\end{equation}

Finally, we introduce the rotating frame which satisfy 
\begin{equation}
	|\Psi(r)\rangle_{\rm R}=e^{iH_0t}|\Psi(r)\rangle_{\rm Schr},
\end{equation}
where $|\Psi(r)\rangle_{\rm R}$ and $|\Psi(r)\rangle_{\rm Schr}$ are states in the rotating frame and the Schr\"odinger picture, respectively, and 
\begin{eqnarray}
H_0&=&(\omega_b+\omega_{\alpha})\sum_{M=-1}^{1}|a,M\rangle_s\langle a,M|+\omega_b\sum_{M=-1}^{1}|b,M\rangle_s\langle b,M|\nonumber\\
&&+(\omega_b
+\omega_{\eta})\sum_{M=-2}^{2}|c,M\rangle_s\langle c,M|.
	\end{eqnarray}
Then the direct calculation yields that in this rotating frame the Hamiltonian is just the one of Eq.~(\ref{hi}).

\begin{acknowledgments}
We appreciate Prof. Yong Li and Prof. Sandra Eibenberger-Arias for helpful discussions.This work is supported by by the Quantum Science and Technology-National Science and Technology Major Project (Grant No.~2023ZD0300700), the National Key Research and Development Program of China (Grant No.~2022YFA1405300), the National Natural Science Foundation of China (Grants No.~12405011), and the Natural Science Foundation of Hainan Province (Grants No.~125QN210).
\end{acknowledgments}

\section*{Data Availability Statement}
The data that support the findings of this study are available within the article. 

%\section*{References}
%\nocite{*}
%\bibliography{ESPref}


\end{document}